\documentclass{moriond}
\usepackage[super]{cite}

\def\be{\begin{equation}}
\def\ee{\end{equation}}
\def\bea{\begin{eqnarray}}
\def\eea{\end{eqnarray}}

\newcommand{\alphas}{\ensuremath{\alpha_{\mathrm{s}}}}

\newcommand{\cfrac}[2]{{\textstyle \frac{#1}{#2}}}
\newcommand{\half}{{\cfrac{1}{2}}}

\newcommand{\mev}{\ensuremath{\hbox{ MeV}}}
\newcommand{\gev}{\ensuremath{\hbox{ GeV}}}
\newcommand{\tev}{\ensuremath{\hbox{ TeV}}}
\newcommand{\ps}{\ensuremath{\hbox{ ps}}}
\newcommand{\ys}{\ensuremath{\hbox{ ys}}}
\newcommand{\fm}{\ensuremath{\hbox{ fm}}}
\newcommand{\fb}{\ensuremath{\hbox{ fb}}}
\newcommand{\pb}{\ensuremath{\hbox{ pb}}}

\newcommand{\jpsi}{\ensuremath{J\!/\!\psi}}

\newcommand{\Photo}{\includegraphics[height=35mm]{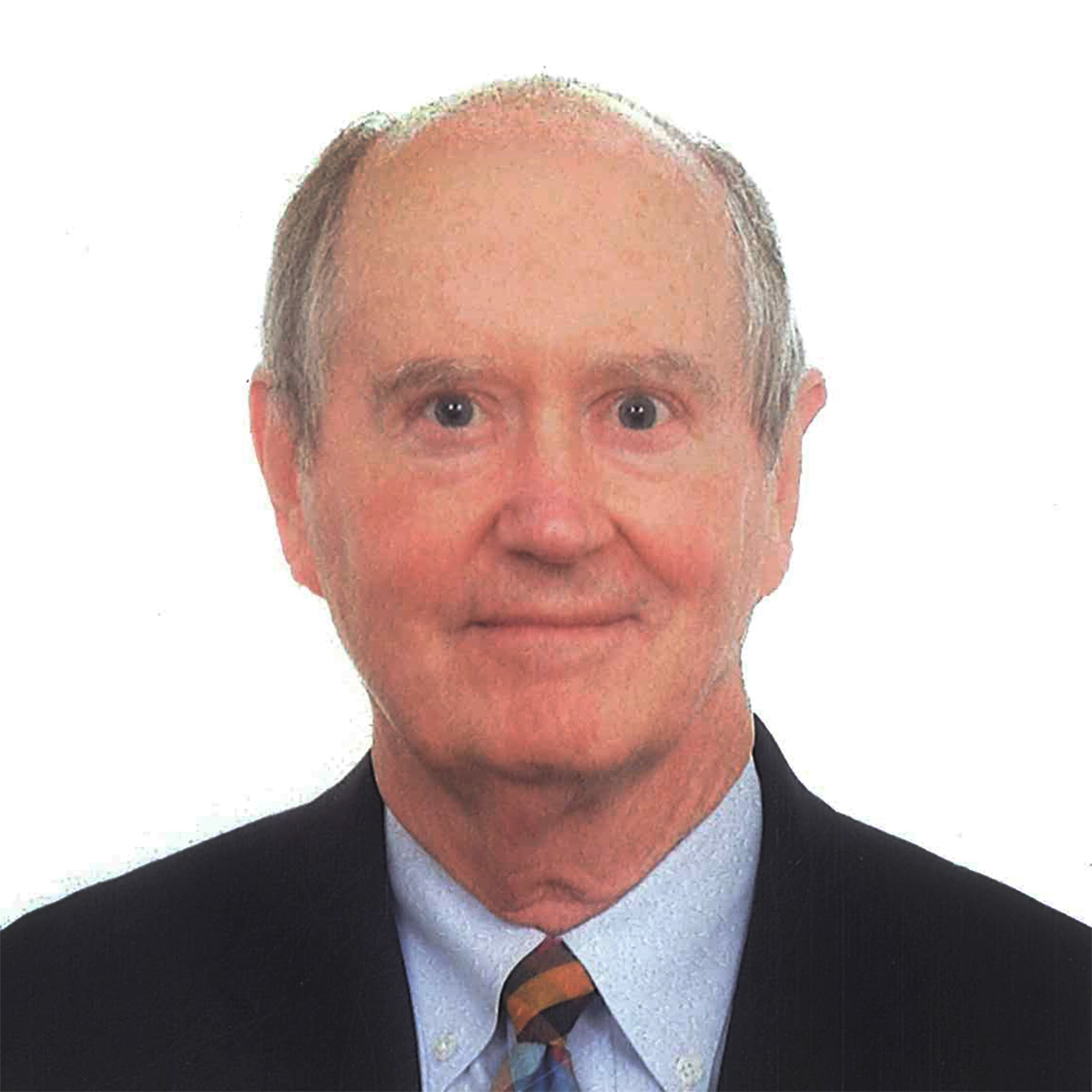}}

\begin{document}
\vspace*{4cm}
\title{STABLE TETRAQUARKS}

\author{ CHRIS QUIGG \footnote{Email: quigg@fnal.gov \hfill \textsf{FERMILAB--CONF--18/099--T}} }

\address{Theoretical Physics Department, Fermi National Accelerator Laboratory \\ P.O.Box 500, Batavia, Illinois 60510 USA}

\maketitle\abstracts{
For very heavy quarks, relations derived from heavy-quark symmetry imply  novel narrow doubly heavy tetraquark states containing two heavy quarks and two light antiquarks. We predict that double-beauty states will be stable against strong decays, whereas the double-charm states and mixed beauty+charm states will dissociate into pairs of heavy-light mesons. Observing  a new double-beauty state through its weak decays would establish the existence of tetraquarks and illuminate the role of heavy color-antitriplet diquarks as hadron constituents.
}

\section{Introduction}
Since the BELLE collaboration's discovery of the charmonium-associated state $X(3872)$,\cite{Choi:2003ue} hadron spectroscopy has been reinvigorated and recast.\cite{Olsen:2017bmm}
Many of the newly observed states invite identification with compositions beyond the traditional quark--antiquark meson and three-quark baryon schemes, possibilities foreseen in the foundational quark-model papers.\cite{Zweig:1981pd} Tetraquark states composed of a heavy quark and antiquark plus a light quark and antiquark have attracted much attention. All the observed candidates fit the form $c \bar c q_k \bar q_l$, where the light quarks $q$ may be $u, d, \hbox{or } s$. 
The  putative tetraquarks typically have strong decays to $c \bar c$ charmonium + light mesons. None is observed significantly below threshold for strong decays into two heavy--light meson states $\bar c q_k + c \bar q_l$.  

Estia Eichten and I have examined the possibility of unconventional tetraquark configurations for which all strong decays are kinematically forbidden.\cite{Eichten:2017ffp} In the heavy-quark limit, stable---hence exceedingly narrow---$Q_iQ_j \bar q_k \bar q_l$ mesons must exist. To apply this insight, we take into account  corrections for finite heavy-quark masses to deduce which tetraquark states containing $b$ or $c$ quarks might be stable. The most promising candidate is a $J^P=1^+$ isoscalar double-$b$ meson, $\mathcal{T}^{\{bb\}-}_{[\bar u \bar d]}$. I will sketch our derivation and results, emphasizing the consequences for experiment, and indicate areas in which experimental and theoretical work can be productive.

\section{Heavy-quark symmetry implies stable heavy tetraquark mesons $Q_iQ_j \bar q_k \bar q_l$}
\begin{figure}
\centerline{\includegraphics[width=0.22\textwidth]{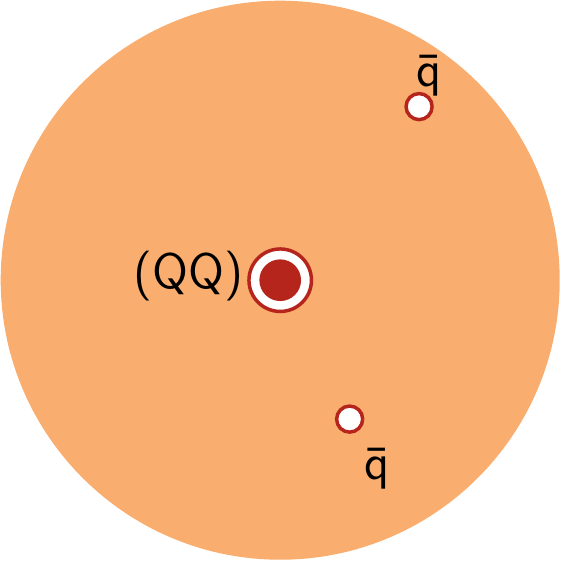}\qquad {\includegraphics[width=0.22\textwidth]{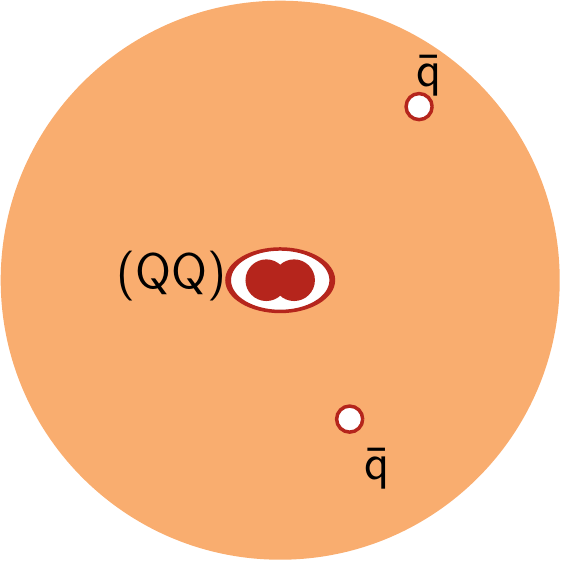}\qquad {\includegraphics[width=0.22\textwidth]{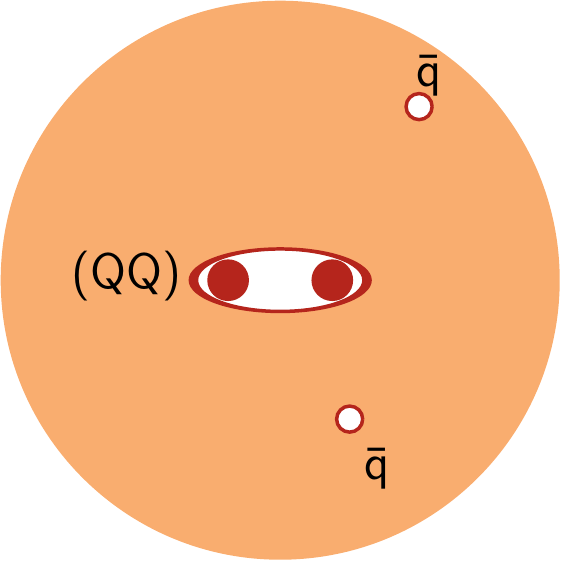}\qquad {\raisebox{-9pt}{\includegraphics[width=0.22\textwidth]{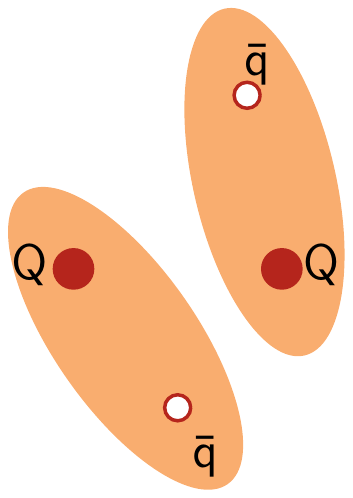}}}}}}
\caption[]{Schematic evolution of  a $Q_iQ_j \bar q_k \bar q_l$ state as the heavy-quark masses decrease (and the mean separation between the heavy quarks increases) from left to right.}
\label{fig:dhtq}
\end{figure}

One-gluon-exchange between a pair of color-triplet heavy quarks is attractive for $(QQ)$ in a color-$\mathbf{\bar{3}}$ configuration and repulsive for the color-$\mathbf{6}$ configuration. The strength of the $\mathbf{\bar{3}}$ attraction is half that of the corresponding $(Q\bar{Q})$ in a color-$\mathbf{1}$. This means that in the limit of very heavy quarks, we may idealize the color-antitriplet $(QQ)$ diquark as a stationary, structureless color charge, as depicted in the leftmost panel in Figure~\ref{fig:dhtq}.
We can separate the strong dynamics binding the diquark from the long-range color interaction by which the light antiquarks interact with each other and are bound to the diquark ``nucleus.''

For sufficiently heavy quarks $Q$, a $Q_iQ_j \bar q_k \bar q_l$ tetraquark meson is stable against strong decays, as we can show by considering possible decay modes. First, we note that dissociation into two heavy--light mesons is kinematically forbidden. The $\mathcal{Q}$ value for the decay is \break
$\mathcal{Q} \equiv m(Q_i Q_j \bar q_k \bar q_l) - [m(Q_i \bar q_k) + m(Q_j \bar q_l)] = 
\Delta(q_k, q_l) - \half\!\left(\cfrac{2}{3}\alphas\right)^2\![1 + O(v^2)]\overline M + O(1/\overline M)$,
where $\Delta(q_k, q_l)$, the contribution due to light dynamics, becomes independent of the heavy-quark masses,  $\overline M \equiv (1/{m_Q}_i + 1/{m_Q}_j)^{-1}$ is the reduced mass of $Q_i$ and $Q_j$, and \alphas\ is the strong coupling. The velocity-dependent hyperfine corrections, here negligible, are calculable in the nonrelativistic QCD formalism.\cite{Caswell:1985ui} For large enough values of $\overline M$, the middle term on the right-hand side dominates, so the tetraquark is stable against decay into two heavy-light mesons.

What of the other possible decay channel, a doubly heavy baryon plus a light antibaryon,
$(Q_iQ_j \bar q_k \bar q_l) \to (Q_iQ_j q_m) + (\bar q_k \bar q_l\bar q_m)$? For very heavy quarks, the contributions of $Q$ motion and spin to the tetraquark mass are negligible.
Since the $(QQ)$ diquark is a color-antitriplet, heavy-quark symmetry tells us that 
$m(Q_iQ_j \bar q_k \bar q_l) - m(Q_iQ_j q_m) = m(Q_x q_k q_l) - m(Q_x \bar q_m)$.
The flavored-baryon--flavored-meson mass difference on the right-hand side has the generic form $\Delta_0 + \Delta_1/{M_Q}_x$. Using the observed mass differences, $m(\Lambda_c) - m(D) = 416.87\mev$ and $m(\Lambda_b) - m(B) = 340.26\mev$, and choosing effective quark masses $m_c \equiv m(\jpsi)/2 = 1.55\gev$, $m_b \equiv m(\Upsilon)/2 = 4.73\gev$, we find $\Delta_1 = 176.6\mev^2$ and $\Delta_0 =303\mev$,  hence the mass difference in the heavy-quark limit is $303\mev$. The right-hand side is in every case smaller than the mass of the lightest antibaryon, $m(\bar p) = 938.27\mev$, so no decay to a doubly heavy baryon and a light antibaryon is kinematically allowed. 

\emph{With no open channels in the heavy-quark limit, stable $Q_iQ_j \bar q_k \bar q_l$ mesons must exist.} To assess the implications for the real world, we must first test whether it makes sense to idealize the $(QQ)$ diquark as a tiny, structureless, color-antitriplet color source.\footnote{See Ref.~\citenum{Richard:2018yrm} for a thoughtful critical assessment.} As the separation between the heavy quarks increases, the light-antiquark cloud screens the $Q_iQ_j$ interaction, altering the $\mathbf{\bar{3}, 6}$ mix, and eventually leading to the division of the $(Q_iQ_j \bar q_k \bar q_l)$ state into a pair of heavy--light mesons. These changes are indicated in the progression from left to right in Figure~\ref{fig:dhtq}. Using a half-strength Coulomb$+$linear quarkonium potential, we verified that the rms core radii are small on the expected tetraquark scale: $\langle r^2\rangle^{1/2} = 0.28\fm\, (cc); 0.24\fm\, (bc); 0.19\fm\, (bb)$. This conclusion is supported by exploratory lattice QCD studies.\cite{Peters:2015tra}

To ascertain whether stable tetraquark mesons might be observed, we must estimate masses of the candidate configurations. Numerous model calculations exist in the literature,\footnote{A useful compilation appears in Table IX of Ref.~\citenum{Luo:2017eub}.} but heavy-quark symmetry makes it possible to compute the $Q_iQ_j \bar q_k \bar q_l$ tetraquark masses directly, through the relation $m(Q_iQ_j \bar q_k \bar q_l) - m(Q_iQ_j q_m) = m(Q_x q_k q_l) - m(Q_x \bar q_m)$, with due attention to spin configurations and finite-mass corrections that arise from hyperfine interactions and kinetic-energy shifts for the light degrees of freedom.\footnote{The arithmetic is made explicit in Ref.~\citenum{Eichten:2017ffp}.} Experiments have determined nearly all the information about heavy baryons and heavy--light mesons needed to evaluate the right-hand side in every case of interest, i.e., for tetraquarks based on $bb, bc, \hbox{and }cc$ diquarks.\footnote{The  lifetime ($\approx 0.4\ys$) of the top quark is too short to permit the formation of hadrons containing $t$.} The doubly heavy baryons have been more elusive: for the moment, the strongest evidence we have is for the $\Xi_{cc}^{++}$ candidate reported by the LHC$b$ experiment at a mass of $3621.40 \pm 0.78\mev$.\cite{Aaij:2017ueg} With this input, we compute the mass of the lightest $(cc)$ tetraquark as $m(\{cc\}[\bar u \bar d]) = 3978\mev$, which lies $102\mev$ above the threshold for decay into $D^+D^{*0}$.\footnote{An earlier sighting by the SELEX Collaboration\cite{Mattson:2002vu} of a $\Xi_{cc}^+$ candidate at $3519\mev$ would imply $m(\{cc\}[\bar u \bar d]) = 3876\mev$, coincident with the threshold for dissociation into a heavy-light pseudoscalar and heavy-light vector. Signatures for weak decay would include $D^+K^-\ell^+\nu$ and $\Xi_c^+\bar n$. The $D^0D^+\gamma$ channel opens at $3734\mev$.} This would be a $J^P = 1^+$ axial-vector meson, symmetric in $cc$ flavor and antisymmetric in the light antiquark flavors. 

In the absence of comprehensive experimental information about the other doubly heavy baryons, we rely for now on model calculations of their masses~\cite{Karliner:2014gca}  as inputs to our tetraquark mass calculation. Our results for the lowest-lying levels are given in Table~\ref{tab:masses}.
\begin{table}[h]
\caption[]{Expectations for ground-state tetraquark masses, in MeV.}
\label{tab:masses}
\vspace{0.4cm}
    \centering
      \begin{tabular}{@{} lcccccc @{}}
      State & $J^P$ &     $m(Q_iQ_j \bar q_k \bar q_l)$  & Example Decay Channel & $\mathcal{Q}$ [MeV] \\
\hline
      $\{cc\}[\bar u \bar d]$ & $1^+$  &    $3978$  & $D^+{D}^{*0}$ 3876 & $102$ \\
      $\{cc\}[\bar q_k \bar s]$ & $1^+$  &    $4156$  & $D^+{D}^{*+}_s$ $3977$ & $179$ \\
      $\{cc\}\{\bar q_k \bar q_l\}$ & $0^+,1^+,2^+$  &    $4146,4167,4210$  & $D^+{D^0}, D^+{D}^{*0}$ $3734, 3876$ & $412, 292, 476$\\
      $[bc][\bar u \bar d]$ & \textcolor{red}{$0^+$}  &    $7229$  & $B^-D^+/B^0D^0$ $7146$& $83$\\
  $[bc][\bar q_k\bar s]$  & $0^+$ &    7406 & $B_s D$ $7236$ & 170 \\     
     $[bc]\{\bar q_k \bar q_l\}$ & $1^+$  &    $7439$  & $B^*D/BD^*$ $7190/7290$ & $249$ \\
 $\{bc\}[\bar u \bar d]$ & $1^+$  &   $7272$  & $B^*D/BD^*$ $7190/7290$& $82$\\
                     $\{bc\}[\bar q_k \bar s]$ & $1^+$ &   7445 &  $ DB_s^*$ 7282 & 163 \\
     $\{bc\}\{\bar q_k \bar q_l\}$ & $0^+,1^+,2^+$  &    $7461,7472,7493$  & $BD/B^*D$ $7146/7190$ & $317,282,349$\\
$\{bb\}[\bar u \bar d]$ & $1^+$  &    $10482$  & $B^-\bar{B}^{*0}$ $10603$& \textcolor{red}{\fbox{$\mathbf{-121}$}} \\
      $\{bb\}[\bar q_k \bar s]$ & $1^+$  &    $10643$  & $\bar{B}\bar{B}_s^*/\bar{B}_s\bar{B}^*$ $10695/10691$ & \textcolor{red}{\fbox{$\mathbf{-48}$}} \\
      $\{bb\}\{\bar q_k \bar q_l\}$ & $0^+,1^+,2^+$  &    $10674,10681,10695$  & $B^-{B^0},B^-{B}^{*0}$ $10559, 10603$ & $115,78, 136$ \\
\hline
   \end{tabular}
\end{table}
We find two real-world candidates for stable tetraquarks: the axial vector $\{bb\}[\bar u \bar d]$  meson, $\mathcal{T}^{\{bb\}-}_{[\bar u \bar d]}$ bound by $121\mev$, and the axial vector $\{bb\}[\bar u \bar s]$ and $\{bb\}[\bar d \bar s]$ mesons bound by $48\mev$. 
Given the provisional doubly heavy baryon masses, we expect all the other $Q_iQ_j \bar q_k \bar q_l$ tetraquarks to lie at least $78\mev$  above the corresponding thresholds for strong decay.\footnote{In model calculations, Karliner and Rosner\cite{Karliner:2017qjm} estimate somewhat deeper binding, and so point to additional $bc$ and $cc$ candidates.} We note that exploratory lattice studies also suggest that double-beauty tetraquarks should be stable.\cite{Bicudo:2015kna,Francis:2016hui} Promising final states  include $\mathcal{T}^{\{bb\}}_{[\bar u \bar d]}(10482)^-\! \to \Xi^0_{bc}\bar{p}$, $B^-D^+\pi^-$, and $B^-D^+\ell^-\bar{\nu}$ (which establishes a weak decay), $\mathcal{T}^{\{bb\}}_{[\bar u \bar s]}(10643)^-\! \to \Xi^0_{bc}\bar{\Sigma}^-$,  $\mathcal{T}^{\{bb\}}_{[\bar d \bar s]}(10643)^0\! \to \Xi^0_{bc}(\bar{\Lambda},\bar{\Sigma}^0)$, and so on. 

If they should lie near enough to threshold, the \emph{unstable} doubly heavy tetraquarks might be observed in ``wrong-sign'' (double flavor) combinations bearing $DD, DB, \hbox{or }BB$ quantum numbers. For example, a $J^P = 1^+\; \mathcal{T}^{\{cc\}}_{[\bar d \bar s]}(4156)^{++} \!\to D^+ D_s^{*+}$ resonance would constitute \emph{prima facie evidence} for a non-$q\bar{q}$ level carrying double charge and double charm. This would be a new kind of resonance, for which no attractive force is present at the meson--meson level. Other nearly bound candidates include $1^+\; \mathcal{T}^{\{bb\}}_{\{\bar q_k \bar q_l\}}(10681)^{0, -, --}$ ($\mathcal{Q} = +78\mev$),  $1^+\; \mathcal{T}^{\{bc\}}_{[\bar u \bar d]}(7272)^0$ ($\mathcal{Q} = +82\mev$), $0^+\; \mathcal{T}^{[bc]}_{[\bar u \bar d]}(7229)^{0}$ ($\mathcal{Q} = +83\mev$), and   $1^+\; \mathcal{T}^{\{cc\}}_{[\bar u \bar d]}(3978)^+$ ($\mathcal{Q} = +102\mev$).

The production of stable doubly heavy tetraquarks (or their nearly bound counterparts) is undoubtedly a rare event, since it entails---at a start---the production of two heavy quarks and two heavy antiquarks. We have no rate calculation to offer, but note the large yield of $B_c$ mesons in the LHC$b$ experiment:\cite{Aaij:2014bva} $8995 \pm 103$ $B_c \to \jpsi\mu\nu_\mu X$ candidates in $2\fb^{-1}$ of $pp$ collisions at $8\tev$, and the CMS observation\cite{Khachatryan:2016ydm} of double-$\Upsilon$ production in 8-TeV $pp$ collisions: $\sigma(pp \to \Upsilon\Upsilon+\hbox{ anything}) = 68 \pm 15\pb$. These suggest that the Large Hadron Collider experiments should be the first focus of searches for novel tetraquark mesons. The ultimate search instrument might be a future electron--positron Tera-$Z$ factory, for which the branching fractions~\cite{Patrignani:2016xqp} $Z \to b\bar{b} =15.12 \pm 0.05\%$ and $Z \to b\bar{b}b\bar{b} = (3.6 \pm 1.3) \times 10^{-4}$ encourage the hope of many events containing multiple heavy quarks.

Two recent investigations go beyond the kinds of arguments I have presented here. Beginning from a situation in which all the constituents are taken to be heavy, so that one-gluon exchange prevails, Czarnecki and collaborators have proposed a figure of merit that governs the color-($\mathbf{\bar{3},6}$) admixture in the putative diquark system.\cite{Czarnecki:2017vco}  They conclude that no stable $QQ\bar{Q}\bar{Q}$ (equal-mass) tetraquarks are to be expected in very-heavy-quark limit, and they find support for the binding of $bb\bar{q}\bar{q}$, in agreement with our conclusions. A generalization allows them to explore how the result depends on $N_\mathrm{c}$, the number of colors. A lattice--NRQCD study of the $bb\bar{b}\bar{b}$ system reveals no tetraquark with mass below $\eta_b\eta_b$, $\eta_b\Upsilon$, $\Upsilon\Upsilon$ thresholds in the $J^{PC} = 0^{++}, 1^{+-}, 2^{++}$ channels.\cite{Hughes:2017xie}

\section{Some tasks to advance our understanding}
\emph{Homework for Experiment.}
The most straightforward request is to \textit{look for double-flavor resonances of two heavy--light mesons near threshold.} The ingredients for such searches should already exist in experiments that have reconstructed many $D$, $D_s$, $B$, and $B_s$ mesons. Next, \textit{extend to $\sqrt{s} = 13\tev$ the measurement of representative cross sections for final states containing two heavy quarks and two heavy antiquarks.} Then we need to \textit{discover and determine the masses of doubly-heavy baryons.} These masses are essential ``engineering information'' for our purposes, as they are needed to implement the heavy-quark--symmetry calculation of tetraquark masses.\footnote{Doubly heavy baryons are of considerable interest in their own right. A light quark bound to a doubly heavy diquark has much in common---in both color configuration and dynamics---with a heavy--light meson. A further goal is to observe excitations of the diquark core, along with the energy levels of the bound light quark.} An important element of the study of doubly heavy baryons is to \textit{resolve the conundrum of the large mass difference between the $\Xi_{cc}^+$ and $\Xi_{cc}^{++}$} candidates reported by SELEX and LHC$b$, respectively. The ultimate experimental goal is to \textit{find stable tetraquarks through their weak decays.}

\emph{Homework for Theory.}
An important challenge is to \textit{develop expectations for the production of final states containing $Q_i, \bar{Q}_i, Q_j, \bar{Q}_j$, and eventually for the anticipated stable tetraquarks.} For the stable $Q_iQ_j \bar q_k \bar q_l$ states we discuss here, \textit{refine lifetime estimates beyond the simplest guess-by-analogy of $\tau \approx 1/3\ps$.}
Extend the considerations of Refs.~\citenum{Richard:2018yrm,Czarnecki:2017vco} to \textit{understand how color configurations evolve with {$QQ$} (and $\bar{q}\bar{q}$) masses.} Continue to explore how diquarks influence hadron spectroscopy, by \textit{analyzing the stability of different body plans in the heavy-quark limit.} A notable example is a possible 
$(Q_iQ_j)(Q_kQ_l)(Q_mQ_n)$ dibaryon, with $\bar{Q}_p\bar{Q}_q\bar{Q}_r$ color structure.


\section{Summary}
In the limit of very heavy quarks $Q$, novel narrow doubly heavy tetraquark states must exist. Heavy-quark symmetry relates the doubly heavy tetraquark mass to the masses of a doubly heavy baryon, heavy-light-light baryon, and heavy-light meson. In the future, when we have more complete experimental knowledge of the doubly heavy baryon spectrum, the heavy-quark--symmetry relations should provide the most reliable predictions of doubly heavy tetraquark masses. Our current mass estimates---which must rely on plausible model inputs for the doubly heavy baryon masses---lead us to expect that the lightest $J^P = 1^+$ $\{bb\}[\bar u \bar d]$, $\{bb\}[\bar u \bar s]$, and $\{bb\}[\bar d \bar s]$ states  should be exceedingly narrow, decaying only through the charged-current weak interaction. The observation of these novel tetraquark mesons would herald a new form of stable matter, in which the doubly heavy color-$\mathbf{\bar 3}$ $(Q_iQ_j)$ diquark is a basic building block. Unstable $Q_iQ_j \bar q_k \bar q_l$ tetraquarks with small $\mathcal{Q}$-values may be observable as resonant pairs of heavy-light mesons in channels with double flavor: $DD, DB, BB$.

\section*{Acknowledgments}
Je souhaite dire un tr\`es grand merci aux gentils organisateurs,
\`a nos amies sauvetrices du secretariat,
au personnel du Planibel,
\`a Kim et Van, et \`a tous les fondateurs des Rencontres de Moriond. 
I am grateful to the Delta Institute for Theoretical Physics and Nikhef, the National Institute for Subatomic Physics, for generous hospitality in Amsterdam, where this note was prepared.
I thank Estia Eichten for collaboration on the work reported here.
This manuscript has been authored by Fermi Research Alliance, LLC under Contract No.\ DE-AC02-07CH11359 with the U.S.\ Department of Energy, Office of Science, Office of High Energy Physics.
%

\end{document}